# Top-Down Approach to Unified Supergravity Models

RALF HEMPFLING

*Deutsches Elektronen-Synchrotron, Notkestraße 85, D-22603 Hamburg, Germany*

## ABSTRACT

We introduce a new approach for studying unified supergravity models. In this approach all the parameters of the grand unified theory (GUT) are fixed by imposing the corresponding number of low energy observables. This determines the remaining particle spectrum whose dependence on the low energy observables can now be investigated. We also include some SUSY threshold corrections that have previously been neglected. In particular the SUSY threshold corrections to the fermion masses can have a significant impact on the Yukawa coupling unification.

# 1. Introduction

It has been shown recently that in the minimal supersymmetric model (MSSM) [1] the $SU(3)_c \times SU(2)_L \times U(1)_Y$ gauge couplings unify at a scale $M_{GUT} = \mathcal{O}(10^{16}$ GeV) [2]. Additionally, the unification of $\tau$ and bottom Yukawa couplings at $M_{GUT}$ can be achieved within the MSSM if the top-Yukawa coupling is close to the Landau-pole [3;4].

The evolution of the coupling constants from the electroweak scale, given by $m_z$, to the GUT scale, $M_{GUT}$, including SUSY threshold corrections is non-trivial without any a priori knowledge of the SUSY particle spectrum. This particle spectrum results from mass parameters that are subject to a renormalization group (RG) evolution from $M_{GUT}$ to $m_z$ and are in general scattered over several orders of magnitude. Several different approaches have been proposed, all of which make some simplified assumptions about the SUSY threshold effects. Also, it is in general not possible to impose all the experimental observables. For example, in the so-called bottom-up approach [5] which is suited to investigate a large number of models $\alpha_s$ is an output rather than an experimental input. Any experimental information on $\alpha_s$ is thus lost.

In general, the goal of all these approaches was to obtain low energy predictions by imposing GUT constraints. In this paper we introduce a complementary approach that enables us to constrain the GUT parameter space by imposing all present and future experimental results. Furthermore, we present a more complete treatment of the SUSY threshold corrections that does not require any assumptions about the SUSY particle spectrum. Our approach is characterized by fixing all the low energy observables including the strong coupling constant, $\alpha_s$ and the bottom mass, $m_b$, precisely. By varying these parameters over the range that is experimentally viable we still obtain the full range of viable SUSY parameters. This way we can explore the possibilities and limitations of probing the GUT parameter space by high-precision measurements. In turn, we can impose GUT constraints in order to study SUSY phenomenology in terms of only a few input parameters.

Our strategy is as follows. We start out with a general $N = 1$ unified supergravity model. In the minimal SU(5) version there are nine input parameters at the GUT scale: the universal gauge coupling, $\alpha_g$, the up and down Yukawa couplings, $\alpha_U$ and $\alpha_D$, the Higgs mass parameter of the superpotential, $\mu$, the GUT scale, $M_{GUT}$, the universal mass parameter of the spin 0 and spin 1/2 particles, $m_0$ and $m_{1/2}$, and the soft SUSY breaking parameters multiplying the trilinear and the quadratic part of the superpotential, $A$ and $B$. (Actually, the Yukawa couplings are $3 \times 3$ matrices, but we can neglect the couplings of the first two generation, which are small and whose origin is unclear. We will also not impose any constraints on the SUSY parameters coming from the non-observation of proton-decay since they are model-dependent.) In order to obtain the low energy parameters we will solve the RG equations (RGEs) twice. At the first run we will determine all the mass parameters and thus all the thresholds at which the RGEs have to be modified. At the second run we will evolve only the coupling constants under consideration of all the SUSY threshold corrections. This way we obtain all the low energy observables including those that are already determined by experiment. By imposing these experimental results we obtain a strongly constrained SUSY particle spectrum whose dependence on the few remaining parameters can now be studied.

Our paper is organized as follows: in Section 2 we describe our approach with all the relevant corrections. In Section 3 we present our numerical results and in Section 4 we summarize our conclusions. The one-loop corrections to the fermion masses are given in the appendix.

## 2. Derivation of the SUSY particle spectrum

The derivation of the SUSY particle spectrum is straightforward. First we evolve all the parameters from $M_{\mathrm{GUT}}$ to $m_z$ using the SUSY $\beta$ functions (two-loop for the couplings [6] and one-loop for the mass parameters [7]) where we impose universality on the mass parameters and unification of the gauge and Yukawa coupling constants. In order to assure that heavy particles with mass $m$ decouple from the set of RGEs at a scale $\sqrt{s} < m$ we multiply the corresponding mass parameter in the $\beta$ function by a step function, $\theta$ [$\theta(x) = 0, 1$ for $x < 0, x \geq 0$, respectively]. Thus we set for the squark, slepton and Higgs mass parameters

$$M_i^2 \to M_i^2 \theta(s - M_i^2), \tag{1}$$

where $i = \widetilde{Q}_n, \widetilde{U}_n, \widetilde{D}_n, \widetilde{L}_n, \widetilde{E}_n, H_m$, ($n = 1, 2, 3; m = 1, 2$), and we decouple the $A$-parameters at scales below the average squark or slepton masses

$$\begin{aligned} A_{U_n} &\to A_{U_n} \theta(2s - M_{\widetilde{Q}_n}^2 - M_{\widetilde{U}_n}^2), \\ A_{D_n} &\to A_{D_n} \theta(2s - M_{\widetilde{Q}_n}^2 - M_{\widetilde{D}_n}^2), \\ A_{E_n} &\to A_{E_n} \theta(2s - M_{\widetilde{L}_n}^2 - M_{\widetilde{E}_n}^2). \end{aligned} \tag{2}$$

For the Higgsino and gaugino mass parameters we have

$$M \to M\theta(s - M^2),  \tag{3}$$

where $M = \mu, M_{\widetilde{B}}, M_{\widetilde{W}}, M_{\tilde{g}}$. With these mass parameters known, we can now compute the full SUSY mass spectrum.

However, the coupling constants we have obtained sofar do not include any threshold corrections. These can be included by evolving the coupling constants a second time while changing the RGEs at every threshold. We start by running the gauge and Yukawa coupling constants, $\alpha_i$ ($i = 1, 2, 3, \tau, b, t$), from $M_{\rm GUT}$ to the scale at which all strongly interacting SUSY particles decouple, $M_{\rm QCD}$, using two-loop SUSY $\beta$ functions. We define $M_{\rm QCD}^2 \equiv M_{\tilde{t}_1} M_{\tilde{t}_2}$ because the decoupling of the top-squark has the strongest effect on the evolution of the Yukawa couplings and the quartic Higgs couplings which below $M_{\rm QCD}$ are allowed to evolve differently than the gauge coupling. The effects of the different squark and gluino thresholds on the gauge couplings, as well as the conversion from DR to $\overline{\rm MS}$ scheme [8] can be easily incorporated at the one-loop level by writing

$$\left(\alpha_i^{-1}\right)_- = \left(\alpha_i^{-1}\right)_+ - \Delta_i,  \tag{4}$$

where the subscript $+$ ($-$) denotes the value above (below) threshold, and we have

$$\Delta_1 = \frac{1}{120\pi} \sum_{i=1,2,3} \left[ \ln\left(\frac{M_{\tilde{Q}_i}^2}{M_{\rm QCD}^2}\right) + 8\ln\left(\frac{M_{\tilde{U}_i}^2}{M_{\rm QCD}^2}\right) + 2\ln\left(\frac{M_{\tilde{D}_i}^2}{M_{\rm QCD}^2}\right) \right],$$

$$\Delta_2 = \frac{1}{8\pi} \sum_{i=1,2,3} \ln\left(\frac{M_{\tilde{Q}_i}^2}{M_{\rm QCD}^2}\right) - \frac{1}{6\pi},$$

$$\Delta_3 = \frac{1}{24\pi} \sum_{i=1,2,3} \left[ \ln\left(\frac{M_{\tilde{D}_i}^2}{M_{\rm QCD}^2}\right) + \ln\left(\frac{M_{\tilde{U}_i}^2}{M_{\rm QCD}^2}\right) + 2\ln\left(\frac{M_{\tilde{Q}_i}^2}{M_{\rm QCD}^2}\right) \right]$$

$$+ \frac{1}{2\pi} \ln\left(\frac{M_{\tilde{g}}^2}{M_{\rm QCD}^2}\right) - \frac{1}{4\pi}.  \tag{5}$$

In the MSSM the Higgs potential is given in general by

$$\mathcal{V} = m_{11}^2 \Phi_1^\dagger \Phi_1 + m_{22}^2 \Phi_2^\dagger \Phi_2 - [m_{12}^2 \Phi_1^\dagger \Phi_2 + {\rm h.c.}]$$

$$+ \tfrac{1}{2}\lambda_1(\Phi_1^\dagger\Phi_1)^2 + \tfrac{1}{2}\lambda_2(\Phi_2^\dagger\Phi_2)^2 + \lambda_3(\Phi_1^\dagger\Phi_1)(\Phi_2^\dagger\Phi_2) + \lambda_4(\Phi_1^\dagger\Phi_2)(\Phi_2^\dagger\Phi_1)$$
$$+ \left\{\tfrac{1}{2}\lambda_5(\Phi_1^\dagger\Phi_2)^2 + \left[\lambda_6(\Phi_1^\dagger\Phi_1) + \lambda_7(\Phi_2^\dagger\Phi_2)\right]\Phi_1^\dagger\Phi_2 + \text{h.c.}\right\}, \quad (6)$$

where the mass parameters are given at $M_{\text{GUT}}$ by $m_{11}^2 = m_{22}^2 = m_0^2 + \mu^2$ and $m_{12}^2 = -\mu B$ and have been evolved down to $M_{\text{QCD}}$. The coupling constants are given at $M_{\text{QCD}}$ by

$$\begin{aligned}
\lambda_1 &= \tfrac{1}{4}(g_2^2 + g_1^2) + \Delta\lambda_1, & \lambda_5 &= \Delta\lambda_5, \\
\lambda_2 &= \tfrac{1}{4}(g_2^2 + g_1^2) + \Delta\lambda_2, & \lambda_6 &= \Delta\lambda_6, \\
\lambda_3 &= \tfrac{1}{4}(g_2^2 - g_1^2) + \Delta\lambda_3, & \lambda_7 &= \Delta\lambda_7, \\
\lambda_4 &= -\tfrac{1}{2}g_2^2 + \Delta\lambda_4.
\end{aligned} \quad (7)$$

Here, the $\Delta\lambda_i$ ($i = 1, .., 7$) are finite one-loop threshold corrections presented in ref. 9. At scales below $M_{\text{QCD}}$ the quartic couplings of the Higgs potential will evolve according to the $\beta$ function of a general two-Higgs-Doublet model (plus sleptons, charginos and neutralinos) [9;10]. With these $\beta$ functions the couplings evolve down to the scale $M_{\text{weak}}^2 \equiv M_{\tilde{e}_1} M_{\tilde{e}_2}$ where all the electroweakly interacting superpartners decouple ($M_{\tilde{e}_1}$ and $M_{\tilde{e}_2}$ denote the selectron mass eigenvalues). We also assume the $m_{A^0} = \mathcal{O}(M_{\text{weak}})$ so that only the standard model (SM) particle content remains at scales below $M_{\text{weak}}$. Thus, for the evolution from $M_{\text{weak}}$ down to $m_z$ we use the SM $\beta$ functions. The SM coupling constants are obtained by imposing the boundary conditions

$$\begin{aligned}
\lambda^{SM} &\equiv \cos^4\beta\,\lambda_1 + \sin^4\beta\,\lambda_2 + 2\cos^2\beta\sin^2\beta\,\widetilde{\lambda}_3 \\
&\quad + 4\cos^3\beta\sin\beta\,\lambda_6 + 4\cos\beta\sin^3\beta\,\lambda_7, \\
\alpha_t^{SM} &= \sin^2\beta\,\alpha_t, \\
\alpha_b^{SM} &= \cos^2\beta\,\alpha_b, \\
\alpha_\tau^{SM} &= \cos^2\beta\,\alpha_\tau.
\end{aligned} \quad (8)$$

where $\widetilde{\lambda}_3 = \lambda_3 + \lambda_4 + \lambda_5$. In order to decouple the top Yukawa coupling from the set of RGEs for $s < m_t$ we write

$$\alpha_t \to \alpha_t \theta(s - m_t^2). \quad (9)$$

Again we use eq. (4) to account for different thresholds, with

$$\Delta_1 = \frac{1}{40\pi}\left\{\ln\left(\frac{m_{A^0}^2}{M_{\text{weak}}^2}\right) + 4\ln\left(\frac{M_{\widetilde{H}}^2}{M_{\text{weak}}^2}\right)\right.$$

$$+ \sum_{i=1,2,3} \left[ \ln \left( \frac{M_{\tilde{L}_i}^2}{M_{\text{weak}}^2} \right) + 2 \ln \left( \frac{M_{\tilde{E}_i}^2}{M_{\text{weak}}^2} \right) \right] \Bigg\} ,$$

$$\Delta_2 = \frac{1}{24\pi} \Bigg\{ 4 \ln \left( \frac{M_{\tilde{H}}^2}{M_{\text{weak}}^2} \right) + 8 \ln \left( \frac{M_{\tilde{W}}^2}{M_{\text{weak}}^2} \right)$$

$$+ \ln \left( \frac{m_{A^0}^2}{M_{\text{weak}}^2} \right) + \sum_{i=1,2,3} \ln \left( \frac{M_{\tilde{L}_i}^2}{M_{\text{weak}}^2} \right) \Bigg\} ,$$

$$\Delta_3 = 0 . \tag{10}$$

where at this point we approximate the CP-odd Higgs mass by $m_{A^0}^2 = m_{11}^2 + m_{22}^2$ and the Higgsino mass by $M_{\tilde{H}} = \mu$. It is understood that all the masses lighter than $m_z$ should be replaced by $m_z$. At the scale $m_z$ we impose the minimum conditions for the Higgs potential of eq. (6)

$$\frac{1}{2}v^2 \cos^2 \beta \left( \lambda_1 + 3\lambda_6 \tan \beta + \tilde{\lambda}_3 \tan^2 \beta + \lambda_7 \tan^3 \beta \right) - \tan \beta m_{12}^2 = 0 ,$$
$$\frac{1}{2}v^2 \sin^2 \beta \left( \lambda_2 + 3\lambda_7 \cot \beta + \tilde{\lambda}_3 \cot^2 \beta + \lambda_6 \cot^3 \beta \right) + m_{22}^2 - \cot \beta m_{12}^2 = 0 . \tag{11}$$

where the low energy effective couplings are given by

$$\lambda_1 = \lambda_1(M_{\text{weak}}) + \cos^4 \beta \Delta \lambda^{SM} ,$$
$$\lambda_2 = \lambda_2(M_{\text{weak}}) + \sin^4 \beta \Delta \lambda^{SM} ,$$
$$\lambda_3 = \lambda_3(M_{\text{weak}}) + \cos^2 \beta \sin^2 \beta \Delta \lambda^{SM} ,$$
$$\lambda_4 = \lambda_4(M_{\text{weak}}) + \cos^2 \beta \sin^2 \beta \Delta \lambda^{SM} ,$$
$$\lambda_5 = \lambda_5(M_{\text{weak}}) + \cos^2 \beta \sin^2 \beta \Delta \lambda^{SM} ,$$
$$\lambda_6 = \lambda_6(M_{\text{weak}}) + \cos^3 \beta \sin \beta \Delta \lambda^{SM} ,$$
$$\lambda_7 = \lambda_7(M_{\text{weak}}) + \sin^3 \beta \cos \beta \Delta \lambda^{SM} , \tag{12}$$

with $\Delta \lambda^{SM} \equiv \lambda^{SM}(m_z) - \lambda^{SM}(M_{\text{weak}})$. This way we have achieved that only the SM Higgs self coupling evolves at scales below $m_{A^0}$. By using running parameters in the Higgs sector we have included all leading log terms summed to all orders in perturbation theory. This formalism yields a better approximation for large SUSY masses [say $\mathcal{O}(1 \text{ TeV})$] than *e.g.* the one-loop effective potential. The top quark threshold corrections to $\alpha_s$ are included via eq. (4) with

$$\Delta_3 = \frac{1}{6\pi} \ln\left(\frac{m_t^2}{m_z^2}\right). \tag{13}$$

Instead of solving eq. (11) for $\tan\beta$ and $v$ we keep $\tan\beta$ fixed and solve for $m_{12}^2$ and $v$. This is possible because until now we have not used the value of $B$ which does not enter the $\beta$ function of any other parameter at one-loop. It can thus be chosen independently of all the other parameters. We now obtain the experimental observables

$$\tan^2\theta_w = \frac{3\alpha_1}{5\alpha_2},$$
$$\alpha_{em} = \sin^2\theta_w \alpha_2,$$
$$\alpha_s = \alpha_3,$$
$$m_z = \frac{\sqrt{\pi\alpha_{em}} v}{\sin\theta_w \cos\theta_w},$$
$$m_t^t = \frac{\sqrt{2\pi\alpha_t} v}{\sin\beta},$$
$$m_b^z = \frac{\sqrt{2\pi\alpha_b} v}{\cos\beta} + \Delta m_b,$$
$$m_\tau^z = \frac{\sqrt{2\pi\alpha_\tau} v}{\cos\beta} + \Delta m_\tau, \tag{14}$$

where the superscripts $t$ and $z$ indicate that these are the running masses evaluated at $m_t$ and $m_z$, respectively. Included are also all the $\tan\beta$ enhanced one-loop SUSY threshold corrections, $\Delta m_b$ and $\Delta m_\tau$, listed in the appendix [11;12]. The top Yukawa coupling is evaluated at $m_t$ and all other couplings at $m_z$. The physical (pole) masses are [13]

$$m_t = m_t^t \left[1 + \frac{4\alpha_s}{3\pi} + 11\left(\frac{\alpha_s}{\pi}\right)^2\right],$$
$$m_b = m_b^b \left[1 + \frac{4\alpha_s}{3\pi} + 12.4\left(\frac{\alpha_s}{\pi}\right)^2\right],$$
$$m_\tau = m_\tau^\tau, \tag{15}$$

We now have to compare out theoretical predictions with experimental data. Our set of experimental quantities is [14]

$$\sin^2\theta_w = 0.2324 - 1.96 \times 10^{-3} \left[\left(\frac{m_t}{138 \text{ GeV}}\right)^2 - 1\right],$$

$$\alpha_{\text{em}}^{-1} = 127.9\,,$$

$$m_z = 91.187 \text{ GeV}\,,$$

$$m_\tau = 1.7841 \text{ GeV}\,. \tag{16}$$

The experimental errors of these quantities are so small that they can be ignored. On the other hand, the QCD gauge coupling carries a significant error. We typically choose $\alpha_s = 0.13$, which is somewhat larger than the result from deep inelastic scattering, $\alpha_s = 0.112$ [15], or even from LEP experiments, $\alpha_s = 0.122$ [16]. This is necessary in order to obtain a SUSY particle spectrum at or below 1 TeV. This discrepancy which is still within the experimental bounds might also be an indication for non-negligible GUT threshold corrections [17]. The bottom mass has been calculated to be $m_b = (4.72 \pm 0.05)$ GeV [18]. We are more conservative and accept $m_b = 5.0$ GeV as the upper limit still compatible with the experimental data. In order to avoid that the discrepancy in $\alpha_s$ propagates into our prediction of $m_b$ (which for $\alpha_s = 0.13$ and $m_t/\sin\beta < 200$ GeV is unacceptably large) we fix in our plots the running bottom mass at $m_z$, $m_b^z = 3.3$ GeV, which corresponds to a physical mass of $m_b = 5.0$ for $\alpha_s = 0.115$. Furthermore, we typically choose $\tan\beta = 35$ and $m_t = 185$ GeV (this corresponds to a running mass of about 172 GeV which is significantly below the IR fixed point [19]).

Now we proceed to fix the GUT input parameters, denoted generically by $I_i$, by imposing the set of experimental observables described above and denoted by $O_j$. Sofar we have described how to obtain these experimental observables as functions of the GUT inputs, $O_j = F_j(I_i)$. However, in practice we need to find $I_i$ as a function of $O_j$. Thus, we have to invert $F_j(I_i)$ numerically. Clearly, this requires the number of inputs to be equal to the number of outputs and can be achieved by defining $I_i$ as the limit of an infinite series

$$I_i = F_i^{-1}(O_j) = \lim_{n\to\infty} I_i^n\,, \tag{17}$$

where we choose a set of initial GUT parameters, $I_i^0$, by default. All the other elements of the series are obtained by iteration

$$I_i^{n+1} = I_i^n + \sum_j (J^n)_{ij}^{-1} \left[O_j - F_j(I_i^n)\right]\,, \tag{18}$$

$(n = 0, 1, ...)$ where we have computed the Jacobian matrices

$$J_{ij}^n \equiv \left.\frac{\partial F_j(I_i)}{\partial I_i}\right|_{I_i=I_i^n}\,, \tag{19}$$

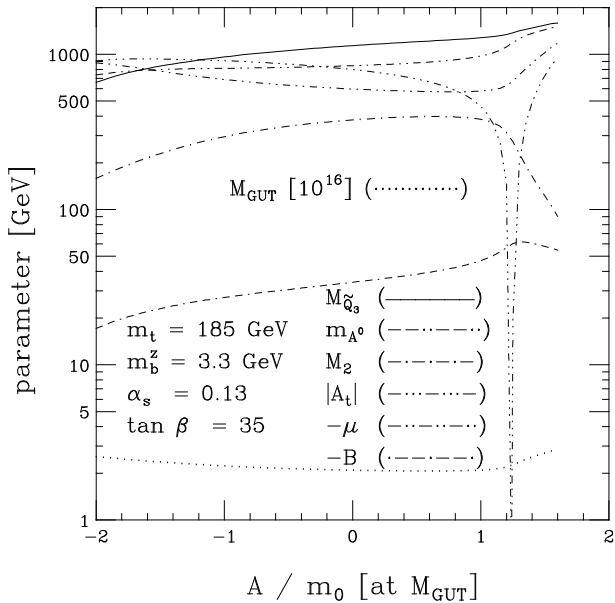

**Fig. 1** The low energy SUSY mass parameters as a function of the $A$-parameter at $M_{\rm GUT}$. The low energy value of $A_t$ is negative unless its initial value at $M_{\rm GUT}$ is very large ($> 1.2 m_0$ in this plot)

and their inverses, $(J^n)_{ij}^{-1}$, numerically. Operationally, we truncate the series when the relative error, $[F_j(I_i^n)/O_j - 1]^2$, is below the desired accuracy. Whether the series defined in eq. (18) converges or not depends strongly on the choice of the initial GUT parameters, $I_i^0$. Thus, the procedure might have to be repeated several times before a set of GUT parameters can be found that yields the desired low energy observables.

## 3. Numerical Results

In the last section we have seen how to obtain the GUT parameters and thus also the full SUSY particle spectrum by imposing low energy observables. We will now present the numerical results of our approach.

### 3.1. Without GUT threshold corrections

Let us assume for now, that the heavy GUT particle spectrum is roughly degenerated and hence that the GUT threshold corrections are negligible. We present our results as functions of $m_b$ and $\alpha_s$ due to their large experimental errors and of $m_t$ and $\tan\beta$, which are still unknown.

A few clarifications are required before presenting the numerical results. In the absence of large SUSY threshold corrections the value of $m_b$ is predicted once we fix $m_\tau$ and $m_t$. Let us now look at the one-loop contributions to $m_b$ [eq. (14)] given in Appendix A. In the limit $M_{\tilde{b}_1} \approx M_{\tilde{b}_2} \gg m_{\rm w}$ we obtain for the dominant gluino and chargino contributions

$$(\Delta m_b)_{\tilde{g}} \approx \frac{\alpha_s}{3\pi} \frac{\mu M_{\tilde{g}}}{m^2} m_b \tan\beta\,,$$
$$(\Delta m_b)_{\chi^\pm} \approx \frac{\alpha_{\rm em}}{16\pi} \frac{m_t^2}{m_{\rm w}^2 s_{\rm w}^2} \frac{\mu A_t}{m^2} m_b \tan\beta\,. \tag{20}$$

where $m = \max\{M_{\tilde{b}_1}, M_{\tilde{b}_2}, M_{\tilde{g}}\}$. Thus, in the large $\tan\beta$ regime (which is under study here) the corrections are strongly enhanced and as a result the bottom mass prediction depends on various other SUSY parameters. Therefore, we can treat $m_\tau$, $m_b$ and $m_t$ as independent parameters.

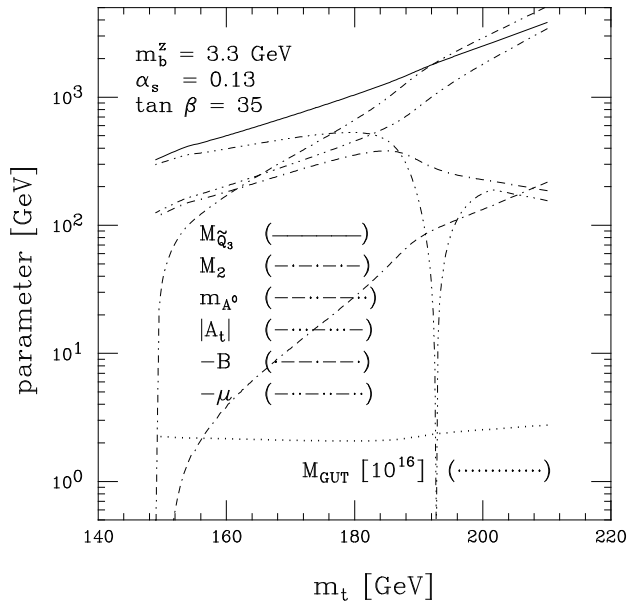

**Fig. 2** The SUSY mass parameters as a function of the top mass. $A_t$ is positive (negative) for $m_t > (<)193$ GeV

It has been shown, that the prediction of $m_b$ from unification without threshold corrections tends to be too large unless the top mass is close to its IR fixed point [3;4]. Thus, we will only obtain acceptable values of $m_b$ for $\mu M_{\tilde{g}} < 0$ [in our plot we will choose $M_{\tilde{g}} > 0$]. By looking at the RGEs we see that $M_{\tilde{g}}$ tends to drive $A_t$ to negative values. As a result, there is a partial cancellation of the terms in eq. (20) over a large portion of the parameter space. Nonetheless, the corrections can be significant.

In fig. 1 we present the SUSY parameters as a function of the $A$-parameter at $M_{\rm GUT}$. We see that the SUSY particle spectrum changes only slowly as long as $|A| \lesssim m_0$. The reason it that the evolution of the $A$ parameters is dominated by $M_{\tilde{g}}$ and thus the low energy value is rather independent of the input at $M_{\text{GUT}}$ which we will choose to be $A = m_0$. In fig. 2–5 we present various low energy SUSY parameters as a function of $m_t$, $m_b$, $\tan\beta$ and $\alpha_s$, respectively. Shown is the region in parameter space where we can find a set of GUT parameters that yields the desired values of the experimental observables. We display the dependence of $M_{\text{GUT}}$, which is important for proton decay, the gaugino and Higgsino mass parameters, $M_2$ and $\mu$, that determine the chargino and neutralino sector, the mass parameter of the left handed top and bottom squark, $M_{\tilde{Q}_3}$, which characterizes the scale of the squark masses and determines the radiative corrections [20] to the lightest Higgs boson mass, $m_{h^0}$, and the mass of the CP-odd scalar

$$m_{A^0}^2 = \frac{2m_{12}^2}{\sin 2\beta} - \tfrac{1}{2}v^2 \left(2\lambda_5 + \lambda_6 \cot\beta + \lambda_7 \tan\beta\right) . \tag{21}$$

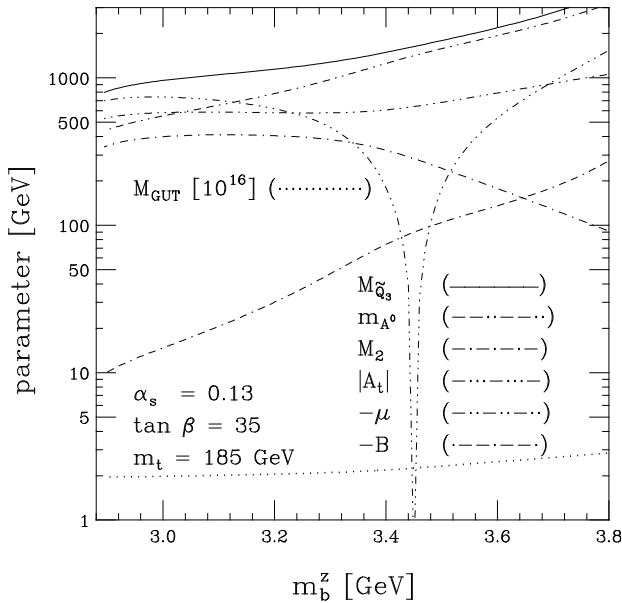

**Fig. 3** The SUSY mass parameters as a function of the running bottom mass, $m_b^z$. The parameter $A_t$ is positive (negative) for $m_b^z > (<)3.45$ GeV.

We also present the $B$-parameter, which turns out to be an important measure for the amount of fine-tuning required to yield large values of $\tan\beta$, and $A_t \equiv A_{U_3}$ for completeness. If one neglects the SUSY threshold corrections [eq. (20)] only a very narrow range of $m_t$ close to the IR fixed point is compatible with $\tau$-bottom Yukawa unification [3;4]. However, in fig. 2 we see that by including these corrections we can find a solution for a much larger range of top masses that is clearly below the IR-fixed point $m_t \approx 200$ GeV [19]. It has been pointed out in ref. 11 that the corrections in eq. (20) are constrained by requiring the absence of severe fine-tuning. The reason is that the minimum conditions of the Higgs potential require that $B\mu/m_0^2 = \mathcal{O}(1/\tan\beta) \ll 1$. Such small values for $B\mu$ are instable under radiative corrections and, therefore, unnatural unless they are protected by an approximate symmetry. Such a symmetry would also require $A_t$ and $m_{1/2}$ to be suppressed which in turn would imply a cancellation of the $\tan\beta$ enhancement of eq. (20). However, our main priority is to look for solutions for a particular set of experimental inputs. Thus, we consider the amount of fine-tuning required for a particular solution as an output which is roughly characterized by the ratio $r \equiv \max\{A_t^2, M_2^2\}/B^2$. The absence of fine-tuning is an important argument in order to find the theoretically favored range in parameter space. For example, the requirement that $r \lesssim 100$ implies that $m_t \gtrsim 180$ GeV. If we ignore the problem of fine-tuning we find a lower limit of $m_t \gtrsim 150$ GeV by requiring that $m_{A^0} > 0$. It is interesting to note that the SUSY particle mass spectrum becomes heavier with $m_t$ as a result of the $m_t$ dependence of $\sin^2\theta_{\rm w}$ [eq. (16)] and of $\alpha_s$ [eq. (13)].

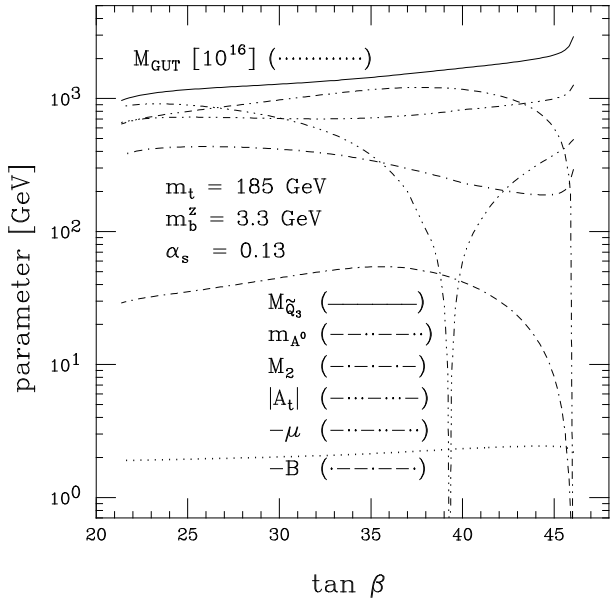

**Fig. 4** The SUSY mass parameters as a function of $\tan\beta$

In fig. 3 we present the bottom mass range for which a solution can be found. It illustrates the significance of the radiative corrections to the $\tau$ and $b$ quark mass [Appendix A]. It is interesting that the $m_b$ dependence shown in fig. 3 exhibits very similar qualitative features to the $m_t$ dependence in fig. 2.

In fig. 4 we show the $\tan\beta$ dependence of the SUSY particle spectrum which is very similar to the $m_t$ and $m_b$ dependence in fig. 2 and 3. The main difference is that the condition $m_{A^0} > 0$ puts an upper limit on $\tan\beta \lesssim h_t/h_b$ [21] as a result of radiative symmetry breaking [22].

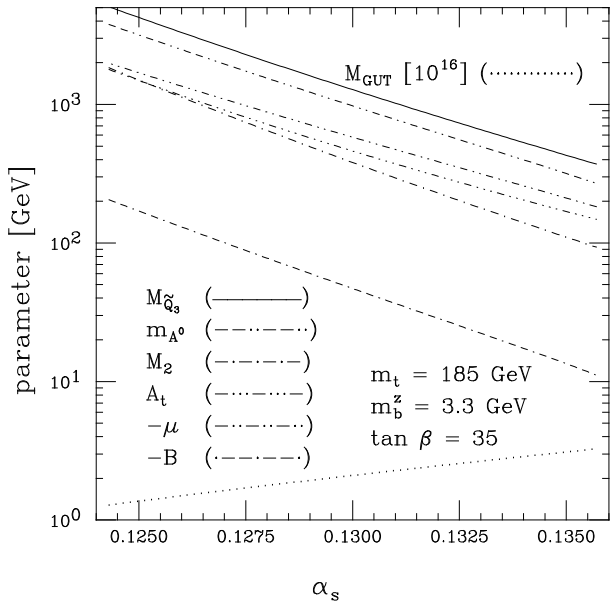

**Fig. 5** The SUSY mass parameters as a function of the strong coupling constant, $\alpha_{rms}$

In fig. 5 we display the range of the strong coupling constants for which gauge and $\tau$-bottom Yukawa unification can be achieved. We find that $M_{\text{GUT}}$ grows with $\alpha_s$ while the SUSY mass spectrum decreases with increasing $\alpha_s$ [2]. However, the actual SUSY particle masses turn out to be much larger than the effective SUSY scale [4]. As a result, the predicted values of $\alpha_s$ from unification are significantly larger than the experimental value as long as the SUSY particle spectrum is not much heavier than 1 TeV. This might be a first indication that the GUT threshold corrections [17] may be significant.

### 3.2. WITH GUT THRESHOLD CORRECTIONS

We now consider the possibility of non-negligible GUT threshold corrections to the gauge couplings. We parameterize these corrections by $\Delta$ such that

$$\alpha_1^{-1} = \alpha_2^{-1} = \alpha_3^{-1} - \Delta = \alpha_g^{-1}. \tag{22}$$

Here, $M_{\text{GUT}}$ is defined as the point at which $\alpha_1$ and $\alpha_2$ intersect. It does in general not correspond to the masses of any heavy particles which are now assumed to be non-degenerate.

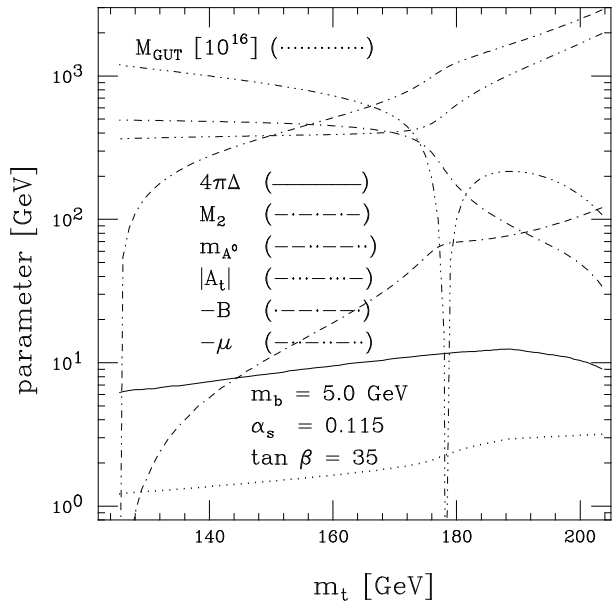

**Fig. 6** The SUSY mass parameters as a function of the top mass. $A_t$ is positive (negative) for $m_t > (<)178$ GeV. Included are finite GUT threshold corrections to the gauge couplings parameterized by $4\pi\Delta$. Instead, we fix the lightest top squark mass, $M_{\tilde{t}_1} = 1$ TeV

In general, we can write

$$4\pi\Delta = \sum_\phi a_\phi \ln\left(\frac{m_\phi^2}{M_{\mathrm{GUT}}^2}\right), \tag{23}$$

where the sum is over all heavy GUT particles, $\phi$, and the coefficients $a_\phi = \mathcal{O}(1)$. For simplicity we set $\alpha_b = \alpha_\tau$ at $M_{\mathrm{GUT}}$. In fig. 6 we present various low energy SUSY parameters as a function of $m_t$. The difference to fig. 2 is that we now fix the lightest top squark mass, $M_{\tilde{t}_1} = 1$ TeV, and instead we allow non-zero values of $\Delta$. This allows us to fix the strong coupling constant at a lower value $\alpha_s = 0.115$ and to fix the physical (pole) bottom mass $m_b = 5.0$ GeV. The solid curve shows the predicted GUT threshold effect defined in eq. (23). We see that the full range of $m_t$ from the experimental lower bound to the IR fixed point is compatible with $\tau$-bottom Yukawa unification within the frame-work of the minimal unified SU(5) SUGRA model.

### 3.3. SUSY LOOP-EFFECTS

One of the main advantages of our top-down approach is that it allows us to greatly reduce the number of free SUSY parameters. Consider, e.g. the mass of the lightest Higgs boson, $m_{h^\circ}$. In the MSSM there exists a well defined tree-level upper limit $m_{h^\circ} \leq m_z$ which implies that the Higgs sector will be a good testing ground for the MSSM even if all SUSY partners are heavy. Through radiative corrections $m_{h^o}$ depends on all SUSY parameters [20] even though not all of them are significant. By imposing GUT relations, universality and the experimental constraints from eq. (16) the number of free parameters can be reduced to five. Furthermore, without great loss of generality we can set $A = m_0$, since the low energy value of $A_t$ is primarily determined by $m_{\tilde{g}}$. In fig. 7 (a) we present the two-loop radiatively corrected prediction of $m_{h^o}$ [20;23] as a function of $m_t$. Here we set $\tan\beta = 35$, $M_{\tilde{t}_1} = 1$ TeV and we have imposed the constraint $\mu = -m_0/2$. This constraint is chosen for convenience since it allows for radiative symmetry breaking over a large range of $m_t$. In fig. 7 (b) we present some physical SUSY particle masses (at tree-level). We see that the constraints $M_{\tilde{\tau}_1}, m_{\tilde{\chi}_1^0}, m_{\tilde{\chi}_1^\pm} \gtrsim m_z/2$ imply 120 GeV $\lesssim m_t \lesssim$ 170 GeV. Of course, this range of $m_t$ for which a solution can be found depends on our initial values of $A$ and $\mu$. However, the value of $m_{h^o}$ as a function will not be affected significantly. Here the QCD coupling and the bottom mass are treaded as outputs and lie in the range $0.122 < \alpha_s < 0.133$ and 5.4 GeV $< m_b <$ 6.6 GeV where the lower (upper) limits correspond to $m_t = 120$ GeV (170 GeV).

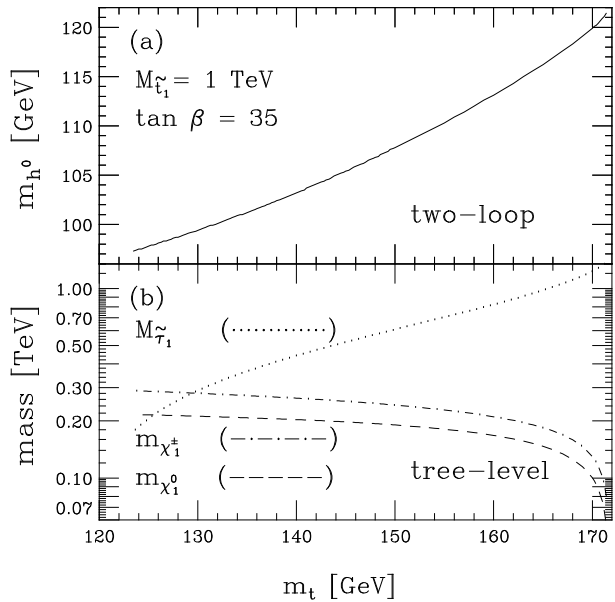

**Fig. 7** The two-loop radiatively corrected lightest Higgs mass (a) and some relevant SUSY particle tree-level masses (b) as a function of $m_t$. We fix $M_{\tilde{t}_1} = 1$ TeV

# 4. Conclusions

We have introduced a top-down approach for studying the GUT parameter space in SUSY-GUT models. With this approach we can analyze explicitly the dependence of the SUSY particle spectrum on various experimental observables. We find that the SUSY particle spectrum becomes heavier with increasing $m_t$ and decreasing $\alpha_s$. Neglecting GUT threshold corrections and assuming all SUSY particles are lighter than 1 TeV we find that $\alpha_s \gtrsim 0.13$. In addition, we have included in our analysis the potentially large SUSY threshold corrections to the quark and lepton masses previously neglected. We focus our attention on the large $\tan\beta$ limit where we have shown that these corrections can in some cases lower the value of $m_t$ for which $\tau$-bottom Yukawa unification is achieved below the present experimental lower limit.

Furthermore, our approach enables us to fix the entire SUSY particle spectrum in terms of a few experimental inputs. This allows the computation of virtual SUSY effects in terms of only a few parameters.

Acknowledgements: I am very grateful to M. Carena and C.E.M. Wagner for sharing their FORTRAN subroutines for the computation of all the $\beta$-functions and for many stimulating conversations. I would also like to thank S. Pokorski and P. Zerwas for many useful discussions.

# APPENDIX

one–loop SUSY threshold corrections to $m_b$ and $m_\tau$

In this appendix we present results for the one-loop radiatively generated down-type fermion masses. The calculation was done in $n$ dimensions using dimensional regularization. We consider the limit of large $\tan\beta$ while keeping the Yukawa couplings $h_u \propto m_u/\sin\beta$ and $h_d \propto m_d/\cos\beta$ constant (here the index $d$ stands for all down type fermions, in particular $b$ and $\tau$). These types of corrections are finite and scheme-independent. The result for the up sfermion-chargino loops is

$$\Delta m_d = -\frac{1}{16\pi^2} \sum_{n,i} M_{\tilde{\chi}_i^\pm} V_{ni}^{L\pm} V_{ni}^{R\pm} B_0(0, M_{\tilde{u}_n}^2, M_{\tilde{\chi}_i^\pm}^2), \qquad (A.1)$$

where $n, i = 1, 2$. The vertices in the basis of electroweak eigenstates are given by [24]

$$\mathcal{V}_{mi}^{L\pm} = \begin{pmatrix} \frac{g m_d U_{i2}}{\sqrt{2} m_w \cos\beta} \\ 0 \end{pmatrix}_m,$$

$$\mathcal{V}_{mi}^{R\pm} = \begin{pmatrix} -gV_{i1}^* \\ \dfrac{gm_u V_{i2}}{\sqrt{2}m_{\rm w}\sin\beta} \end{pmatrix}_m . \qquad (A.2)$$

The chargino mass eigenvalues, $M_{\tilde{\chi}_i^\pm}$, and rotation matrices, $U$ and $V$, are defined via

$$\mathrm{diag}(M_{\tilde{\chi}_1^\pm}, M_{\tilde{\chi}_2^\pm}) = UXV^{-1}, \quad X = \begin{pmatrix} M & m_{\rm w}\sqrt{2} \\ 0 & -\mu \end{pmatrix}, \qquad (A.3)$$

From here we obtain the vertices for the mass eigenstates by rotating

$$V_{ni}^{P\pm} = \mathcal{U}(\theta_{\tilde{u}})_{nm}\mathcal{V}_{mi}^{P\pm}, \qquad (A.4)$$

where $P = L, R$ and the mixing angle and the unitary matrix are defined by

$$\sin 2\theta_{\tilde{u}} \equiv \frac{2(A_u - \mu\cot\beta)m_u}{m_{\tilde{u}_1}^2 - m_{\tilde{u}_2}^2},$$
$$\mathcal{U}(\theta) = \begin{pmatrix} \cos\theta & \sin\theta \\ -\sin\theta & \cos\theta \end{pmatrix}. \qquad (A.5)$$

Within the framework of a large $\tan\beta$ approximation we set $m_d = 0$ for consistency and finiteness. In this case, the conventionally defined scalar two-point function becomes

$$B_0(0, m_1^2, m_2^2) = \Delta - \frac{m_1^2 \ln m_1^2 - m_2^2 \ln m_2^2}{m_1^2 - m_2^2} + 1, \qquad (A.6)$$

where $\Delta = 2/(4-n) - \gamma_{\rm E} + \ln(4\pi)$ and $\gamma_{\rm E}$ is the Euler constant. The result for the down sfermion-neutralino loops is

$$\Delta m_d = -\frac{1}{16\pi^2}\sum_{i,n} m_{\chi_i^0} V_{ni}^{L0} V_{ni}^{R0} B_0(0, M_{\tilde{d}_n}^2, M_{\tilde{\chi}_i^0}^2), \qquad (A.7)$$

where $i = 1,2,3,4$ and $n = 1,2$. The vertices for the mass eigenstates are obtained by rotation

$$V_{ni}^{P0} = \mathcal{O}(\theta_{\tilde{d}})_{nm}\mathcal{V}_{mi}^{P0},$$
$$\sin 2\theta_{\tilde{d}} \equiv \frac{2(A_d - \mu\tan\beta)m_d}{m_{\tilde{d}_1}^2 - m_{\tilde{d}_2}^2}, \qquad (A.8)$$

where we have defined

$$\mathcal{V}_{mi}^{L0} = \frac{1}{\sqrt{2}} \begin{pmatrix} 2ee_d Z'_{i1} - \frac{g}{c_{\mathrm{w}}}\left(1 + 2e_d s_{\mathrm{w}}^2\right) Z'_{i2} \\ \frac{g m_d}{m_{\mathrm{w}} \cos\beta} Z_{i3} \end{pmatrix}_m ,$$

$$\mathcal{V}_{mi}^{R0} = \frac{1}{\sqrt{2}} \begin{pmatrix} \frac{g m_d}{m_{\mathrm{w}} \cos\beta} Z_{i3} \\ -2ee_d Z'_{i1} + 2\frac{g}{c_{\mathrm{w}}} e_d s_{\mathrm{w}}^2 Z'_{i2} \end{pmatrix}_m . \qquad (A.9)$$

Here the neutralino mass eigenvalues, $M_{\tilde{\chi}_i^0}$, and rotation matrix, $Z$, are defined via

$$\mathrm{diag}(M_{\tilde{\chi}_1^0}, M_{\tilde{\chi}_2^0}, M_{\tilde{\chi}_3^0}, M_{\tilde{\chi}_4^0}) = ZYZ^{-1} ,$$

$$Y = \begin{pmatrix} M' & 0 & 0 & m_{\mathrm{z}} s_{\mathrm{w}} \\ 0 & M & 0 & -m_{\mathrm{z}} c_{\mathrm{w}} \\ 0 & 0 & 0 & -\mu \\ m_{\mathrm{z}} s_{\mathrm{w}} & -m_{\mathrm{z}} c_{\mathrm{w}} & -\mu & 0 \end{pmatrix} , \qquad (A.10)$$

The result for the down squark-gluino loop is (these type of corrections are absent for the leptons)

$$\Delta m_d = \frac{\alpha_{\mathrm{s}}}{3\pi} M_{\tilde{g}} \sin 2\theta_{\tilde{d}} \\ \times \left[ B_0(0, M_{\tilde{d}_1}^2, M_{\tilde{g}}^2) - B_0(0, M_{\tilde{d}_2}^2, M_{\tilde{g}}^2) \right] . \qquad (A.11)$$